\documentstyle[aps,psfig,floats]{revtex}
\begin{document}
\draft
\wideabs{

\title{Magnetoresistance of $\bf YBa_2Cu_3O_7$ in the ``cold spots'' model}

\author{Anatoley T. Zheleznyak\cite{Zheleznyak}, Victor
M. Yakovenko\cite{Yakovenko}, and H. D. Drew\cite{Drew}}
\address{Department of Physics and Center for Superconductivity
Research, University of Maryland, College Park, Maryland 20742}
\date{\bf cond-mat/9807058, v.1: July 3, 1998, v.2: August 17, 1998}
\maketitle

\begin{abstract}%
We calculate the in-plane magnetoresistance
$\Delta\rho_{xx}/\rho_{xx}$ of YBa$_2$Cu$_3$O$_7$ in a magnetic field
applied perpendicular to the $\rm CuO_2$ planes for the ``cold spots''
model.  In this model, the electron relaxation time
$\tau_2\propto1/T^2$ at small regions on the Fermi surface near the
Brillouin zone diagonals is much longer than the relaxation time
$\tau_1\propto1/T$ at the rest of the Fermi surface ($T$ is
temperature).  In qualitative agreement with the experiment, we find
that Kohler's rule is strongly violated, but the ratio
$\Delta\rho_{xx}/\rho_{xx}\tan^2\theta_H$, where $\tan\theta_H$ is the
Hall angle, is approximately temperature-independent.  We find the
ratio is about 5.5, which is of the same order of magnitude as in the
experiment.
\end{abstract}

}

\begin{figure}
\centerline{\psfig{file=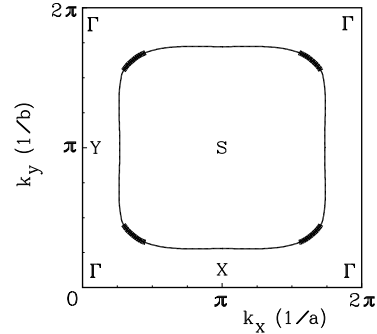,width=0.7\linewidth,angle=-90}}
\caption{Fermi surface of the ${\rm CuO_2}$ bonding band of ${\rm
YBa_2Cu_3O_7}$ according to the band-structure calculations
\protect\cite{Mazin94}.  The thick lines indicate the ``cold spots''.}
\label{fig:FS}
\end{figure}

Theoretical interpretation of the electron transport properties in the
normal state of high-temperature superconductors has attracted a great
deal of attention (see references and discussion in Ref.\
\cite{Yakovenko98a}).  In Ref.\ \cite{Yakovenko98a}, we found that the
frequency dependences of the longitudinal $\sigma_{xx}(\omega)$ and
Hall $\sigma_{xy}(\omega)$ conductivities, measured in
YBa$_2$Cu$_3$O$_7$ in Ref.\ \cite{Drew96}, can be adequately fitted
within a phenomenological model where different regions of the Fermi
surface are characterized by two different relaxations times $\tau_1$
and $\tau_2$.  We called this model the additive two-$\tau$ model.  We
found that the electron relaxation time $\tau_2$ at the ``cold
spots'', small regions on the Fermi surface near the Brillouin zone
diagonals shown by the thick lines in Fig.\ \ref{fig:FS}, is much
longer than the relaxation time $\tau_1$ at the rest of the Fermi
surface: $\tau_2/\tau_1\approx 4$ at $T=95$ K \cite{Yakovenko98a}.
The microscopic origin of the cold spots is not very clear, but their
pattern suggests that they may be related to the $d$-wave symmetry of
the pseudogap in the normal state of cuprates \cite{Ioffe98}.  In
Ref.\ \cite{Yakovenko98a}, we assumed for simplicity that the electron
relaxation time changes discontinuously between the values $\tau_1$
and $\tau_2$ as a function of position on the Fermi surface.
Recently, another version of the cold spots model was independently
introduced by Ioffe and Millis \cite{Ioffe98}.  In their model, the
electron transport properties are completely determined by the cold
spots, where the electron relaxation rate $1/\tau$ has a deep minimum:
\begin{equation}
  1/\tau(k_t)-1/\tau(\bar k_t)\propto(k_t-\bar k_t)^2.
\label{eq:k^2}
\end{equation}
In Eq.\ (\ref{eq:k^2}), $k_t$ is the component of the electron wave
vector tangential to the Fermi surface, which labels different points
on the Fermi surface, and $\bar k_t$ is its value at the center of a
cold spot.  Ioffe and Millis noticed that their model produces a
magnetoresistance too high to agree with the experiment \cite{Ong95}.
Magnetoresistance was discussed theoretically for the spinon-holon
model in Ref.\ \cite{Ong95}, for the change-conjugation model in Ref.\
\cite{Boebinger98}, and for the nearly-antiferromagnetic Fermi liquid
model in Ref.\ \cite{Pines98}.  In the present paper, we evaluate
magnetoresistance for the additive two-$\tau$ model of Ref.\
\cite{Yakovenko98a}. (See also a comment on this model in Ref.\
\cite{Khveshchenko98}.)

The high-temperature cuprate superconductors can be modeled as
two-dimensional (2D) square lattices parallel to the $(x,y)$ plane and
spaced with the distance $d$ along the $z$ axis. We consider the case
where a weak magnetic field $H$ is applied perpendicular to the planes
and study the in-plane transport properties.  For this purpose, we may
neglect coupling between the layers and consider the 2D electron wave
vector {\bf k} and energy $\varepsilon({\bf k})$.  The linearized
stationary Boltzmann equation \cite{Abrikosov}, written for the
deviation $\chi({\bf k})$ of the electron distribution function
$f({\bf k})$ from the equilibrium Fermi function $f_0(\varepsilon)$:
$f({\bf k})-f_0(\varepsilon)= -\chi({\bf k})\,\partial
f_0(\varepsilon)/\partial\varepsilon$, has the form:
\begin{equation}
  \frac{eH}{\hbar c}\,l(k_t)\,\frac{d\chi(k_t)}{d k_t}
  +\chi(k_t)=e{\bf E}\cdot{\bf l}(k_t),
\label{eq:chi}
\end{equation}
where ${\bf l}(k_t)={\bf v}(k_t)\tau(k_t)$ and ${\bf
v}(k_t)=\partial\varepsilon({\bf k})/\hbar\partial{\bf k}$ are the
local values of the electron mean-free path and velocity at a point
$k_t$ on the Fermi surface. ($e$ is the electron charge, $\hbar$ is
Planck's constant, and $c$ is the speed of light.)

In the case of a weak magnetic field $H$, the first term in Eq.\
(\ref{eq:chi}) is a small perturbation, so a solution can be obtained
as the Jones-Zener series in powers of $H$ by iterating this term:
\begin{equation}
  \chi(k_t)=e\sum_{n=0}^{\infty}
  \left(-\frac{eH}{\hbar c}\,l(k_t)\,\frac{d}{d k_t}\right)^n
  {\bf E} \cdot {\bf l}(k_t).
\label{chi_it}
\end{equation} 
Using Eq.\ (\ref{chi_it}), we calculate the electric current ${\bf j}
= 2e \sum_{\bf k} {\bf v}({\bf k}) f({\bf k})$, where the coefficient
2 accounts for two spin orientations of electrons, and find the
conductivity tensor:
$$
  \sigma_{ij}=\frac{2e^2}{(2\pi)^2 \hbar d} \sum_{n=0}^{\infty}
  \oint dk_t \frac{l_i(k_t)}{l(k_t)}
  \left(- \frac{eH}{\hbar c}\,l(k_t)\,\frac{d}{d k_t}\right)^n l_j(k_t).
$$
Here the integral in $k_t$ is taken around the Fermi surface.  For a
system of tetragonal symmetry, the diagonal conductivity
$\sigma_{xx}=\sigma_{xx}^{(0)}+\sigma_{xx}^{(2)}$ up to the second
order in $H$ and the Hall conductivity $\sigma_{xy}$ linear in $H$ are
given by the following expressions \cite{Ong91a,Ong95}:
\begin{eqnarray}
  \sigma_{xx}^{(0)} &=&\frac{e^2}{(2\pi)^2 \hbar d} \oint dk_t\, l(k_t),
\label{sxx0} \\
  \sigma_{xy} &=& \frac{e^3H}{(2\pi)^2 \hbar^2 c d} 
  \oint {\bf e}_z \cdot \left[ {\bf l}(k_t) \times d{\bf l}(k_t) \right],
\label{sxy} \\
  \sigma_{xx}^{(2)} &=& \frac{2e^4 H^2}{(2\pi)^2 \hbar^3 c^2 d} 
  \oint dk_t\, l_x(k_t) \, \frac{d}{dk_t} 
  \left( l(k_t) \frac{dl_x(k_t)}{dk_t}\right),
\nonumber \\
  &=&-\frac{2e^4 H^2}{(2\pi)^2 \hbar^3 c^2 d} \oint dk_t\, l(k_t) 
  \left(\frac{dl_x(k_t)}{dk_t}\right)^2,
\label{sxx2}
\end{eqnarray}
where ${\bf e}_z$ is a unit vector along the $z$ axis. According to
Eq.\ (\ref{sxx2}), magnetic field always reduces the conductivity:
$\sigma_{xx}^{(2)}<0$.

When evaluated for a discontinuous function $\tau(k_t)$, Eqs.\
(\ref{sxx0}) and (\ref{sxy}) give finite results, which were studied
in Ref.\ \cite{Yakovenko98a}.  However, Eq.\ (\ref{sxx2}) gives an
infinite value for $\sigma_{xx}^{(2)}$ because of the squared
derivative $[d\tau(k_t)/dk_t]^2$ of a discontinuous function
$\tau(k_t)$.  In order to obtain a finite result, we introduce a
smooth interpolation between the two values of $\tau$ in a transition
interval of the width $\kappa$:
\begin{equation}
  \tau(k_t) = \frac{\tau_1+\tau_2}{2} \mp \frac{\tau_1+\tau_2}{2} 
  \tanh \left(\frac{k_t-\tilde{k}_t}{\kappa}\right),
\label{tau_trans}
\end{equation}
where $\tilde{k}_t$ is the boundary between the ``hot'' and ``cold''
regions.  We find the following three contributions to
$\sigma_{xx}^{(2)}$:
\begin{equation}
  \sigma_{xx}^{(2)}=-C_1\tau_1^3 -C_2\tau_2^3-\tilde\sigma_{xx}^{(2)}.
\label{sxx2p}
\end{equation}
The coefficients $C_1$ and $C_2$ are given by the integrals over the
``hot'' and ``cold'' regions of the Fermi surface, where the
relaxation times are $\tau_1$ and $\tau_2$, respectively:
\begin{equation}
  C_{1,2}=\frac{2e^4 H^2}{(2\pi)^2 \hbar^3 c^2 d} \int_{1,2} dk_t\, v(k_t) 
  \left(\frac{dv_x(k_t)}{dk_t}\right)^2.
\label{eq:C12}
\end{equation}
The first two terms in Eq.\ (\ref{eq:C12}) do not depend on $\kappa$.
The last term in Eq.\ (\ref{sxx2p}) comes from the transition interval
between the ``hot'' and ``cold'' areas:
\begin{equation}
  \tilde\sigma_{xx}^{(2)} = \frac{2e^4 H^2}{(2\pi)^2 \hbar^3 c^2 d}\,
  v^3({\tilde k}_t)\,\frac{(\tau_1+\tau_2)(\tau_1-\tau_2)^2}{6\kappa}.
\label{sxx2_sing}
\end{equation}
This term becomes singular when the width of the transition interval
is reduced to zero: $\tilde\sigma_{xx}^{(2)}\to\infty$ when
$\kappa\to0$.

Using Eqs.\ (\ref{sxx2p}), (\ref{eq:C12}), and (\ref{sxx2_sing}) with
the parameters determined in Ref.\ \cite{Yakovenko98a}, we calculate
$\sigma_{xx}^{(2)}$ for the $\rm CuO_2$ bonding band of $\rm
YBa_2Cu_3O_7$ shown in Fig.\ \ref{fig:FS}.  As in Ref.\
\cite{Yakovenko98a}, we neglect contributions of the other bands.  We
phenomenologically assign linear and quadratic temperature dependences
to the scattering rates in the ``hot'' and ``cold'' regions:
\begin{equation}
  \tau_1^{-1}\propto T, \qquad \tau_2^{-1}\propto T^2,
\label{eq:T12}
\end{equation}
with the same parameters as Ref.\ \cite{Yakovenko98a}.  Computing the
integrals (\ref{eq:C12}) over the Fermi surface, we find that the
ratio $C_1:C_2$ turns out to be the same as the ratio
$b_1:b_2=0.71:0.29$ in Ref.\ \cite{Yakovenko98a}.

\begin{figure}
\centerline{\psfig{file=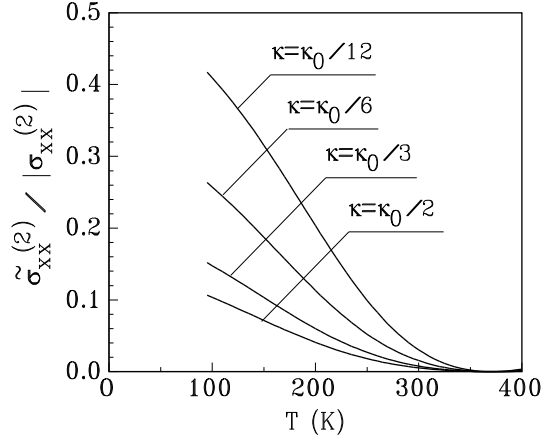,width=\linewidth,angle=-90}}
\caption{Contribution of the transitory zone between the ``hot'' and
  ``cold'' regions on the Fermi surface to magnetoconductivity,
  $\tilde\sigma_{xx}^{(2)}$, normalized to the total
  magnetoconductivity $|\sigma_{xx}^{(2)}|$ and shown as a function of
  temperature.  Different curves correspond to different sizes of the
  transition interval $\kappa$, indicated as a fraction of the size
  $\kappa_0$ of a ``cold spot''.}
\label{fig:ratio}
\end{figure}

The size of the transition interval, $\kappa$, is limited from above
by the size $\kappa_0=0.6/a$ of a cold spot: $\kappa\leq\kappa_0$,
where $a$ is the $\rm CuO_2$ lattice spacing.  In Fig.\
\ref{fig:ratio}, we show the ratio
$\tilde\sigma_{xx}^{(2)}/|\sigma_{xx}^{(2)}|$ computed for different
values of $\kappa$.  We observe that $\tilde\sigma_{xx}^{(2)}$
produces a significant contribution to $\sigma_{xx}^{(2)}$ only when
the transition interval becomes very small: $\kappa\le\kappa_0/12$.
For a realistic width, $\kappa=\kappa_0/2$, the contribution of the
transition interval is insignificant:
$\tilde\sigma_{xx}^{(2)}\ll|\sigma_{xx}^{(2)}|$.

\begin{figure}
\centerline{\psfig{file=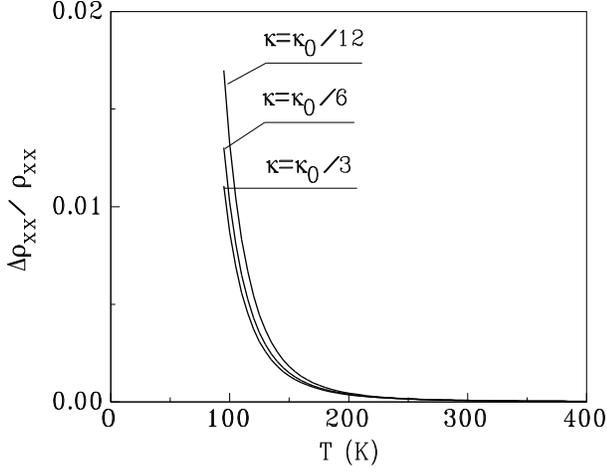,width=\linewidth,angle=-90}}
\caption{Magnetoresistance $\Delta\rho_{xx}/\rho_{xx}$ as a function
  of temperature in the magnetic field $H=9$ T.}
\label{fig:MR}
\end{figure}

The experimentally measurable quantity is the magnetoresistance
$\Delta \rho_{xx}=\rho_{xx}(H,T)-\rho_{xx}(0,T)$, which is related to
the conductivities by the formula
\begin{equation}
  \Delta\rho_{xx}/\rho_{xx} = -\sigma_{xx}^{(2)}/\sigma_{xx}^{(0)} 
  - \tan^2\theta_H,
\label{eq:MR}
\end{equation}
where $\tan\theta_H = \sigma_{xy}/\sigma_{xx}^{(0)}$ is the Hall
angle.  Fig.\ \ref{fig:MR} shows the calculated temperature dependence
of magnetoresistance $\Delta\rho_{xx}/\rho_{xx}$ at $H=9$ T for
different values of $\kappa$.  

It is more convenient to study the dimensionless ratio
\begin{equation}
  \zeta=\Delta\rho_{xx}/\rho_{xx}\tan^2\theta_H
  = -\sigma_{xx}^{(2)}\sigma_{xx}^{(0)}/\sigma_{xy}^2 -1,
\label{eq:zeta}
\end{equation}
which does not depend on magnetic field. In the Drude model with a
single relaxation time $\tau(T)$, $\zeta$ does not depend on
temperature, because $\tau(T)$ cancels out, and is equal to
\begin{equation}
  \bar\zeta=\frac{\oint dk_t\,v(k_t)(dv_x(k_t)/dk_t)^2\oint
  dk_t\,v(k_t)} {2(\oint dk_t\, v_x(k_t)\,dv_y(k_t)/dk_t)^2} - 1.
\label{eq:zeta_bar}
\end{equation}
It was found experimentally that $\zeta$ is nearly
temperature-independent and equals $1.5\div1.7$ in $\rm YBa_2Cu_3O_7$,
13.6 in $\rm La_2Sr_xCuO_4$ \cite{Ong95}, 3.6 in optimally doped and
2.0 in overdoped $\rm Tl_2Ba_2CuO_{6+\delta}$
\cite{Boebinger98,Tyler}.  These results are quite remarkable given
that electron transport in cuprates {\it cannot} be described by a
single-relaxation-time Drude model.  In the spinon-holon model of
Ref.\ \cite{Ong95}, $\zeta$ is given by the same
temperature-independent expression (\ref{eq:zeta_bar}) as in the
simple Drude model, because the two relaxation times $\tau_{\rm tr}$
and $\tau_H$ cancel out.

\begin{figure}
\centerline{\psfig{file=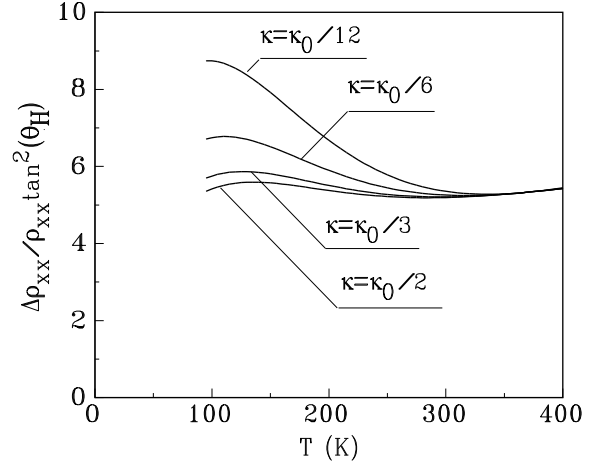,width=\linewidth,angle=-90}}
\caption{Dimensionless ratio of magnetoresistance and the Hall angle
  squared, $\Delta\rho_{xx}/\rho_{xx}\tan^2\theta_H$, as a function of
  temperature for different values of $\kappa$.}
\label{fig:MRrHA}
\end{figure}

In Fig.\ \ref{fig:MRrHA}, we plot $\zeta(T)$ calculated using Eqs.\
(\ref{sxx2p})--(\ref{eq:zeta}) for different values of $\kappa$.  For
a realistic $\kappa=\kappa_0/2$, we observe that $\zeta$ is
approximately temperature-independent.  This is quite surprising
considering that $\zeta(T)$ is given by a complicated rational
function of temperature.  This result demonstrates that, contrary to
the conclusion made in Ref.\ \cite{Ong95}, $\zeta$ can be
temperature-independent in a Fermi-liquid model
\cite{Yakovenko98a,Cooper92,Mihaly92}, where different regions on the
Fermi surface have different temperature dependences of $\tau$ given
by Eq.\ (\ref{eq:T12}).  We wish to emphasize that we do not optimize
the parameters of the model to make $\zeta$ temperature-independent;
we use the same parameters that have been found already in Ref.\
\cite{Yakovenko98a} by fitting frequency dependences of $\sigma_{xx}$
and $\sigma_{xy}$.  As Fig.\ \ref{fig:MRrHA} shows, the value of
$\zeta$ is about 5.5, which is close to the Drude value
$\bar\zeta=5.67$ calculated from Eq.\ (\ref{eq:zeta_bar}) for the
Fermi surface shown in Fig.\ \ref{fig:FS}.  While the theoretical
value of $\zeta$ is greater than the experimental one in $\rm
YBa_2Cu_3O_7$ by a factor of 3, both values are of the same order of
magnitude.  The discrepancy could be ascribed to an excessive
flatness of our model Fermi surface compared with the real one, which
reduces $\tan^2\theta_H$ and increases $\zeta$.  This suggestion is
supported by the fact that the highest value of $\zeta$ is found in
$\rm La_2Sr_xCuO_4$, where the Fermi surface is believed to be the
flattest. (Eq.\ (\ref{eq:zeta_bar}) gives zero for a circular Fermi
surface.)

\begin{figure}
\centerline{\psfig{file=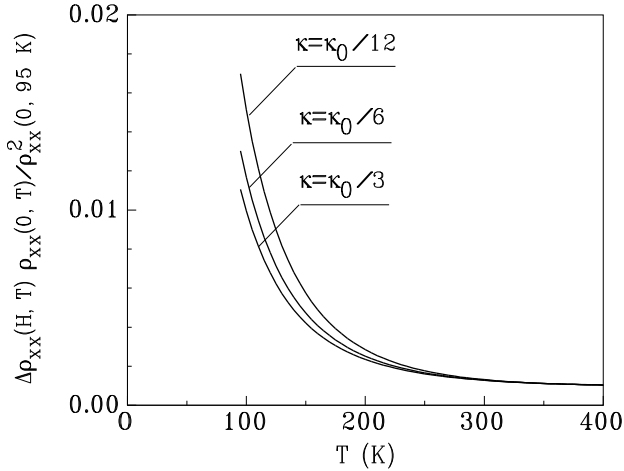,width=\linewidth,angle=-90}}
\caption{The combination
  $\Delta\rho_{xx}(H,T)\rho_{xx}(0,T)/\rho_{xx}^2(0,\mbox{95 K})$ that
  would be temperature-independent if Kohler's rule were satisfied.
  The curves are calculated for the magnetic field $H=9$ T.}
\label{fig:Kohler}
\end{figure}

Because $\zeta$ is approximately temperature-independent, and our
model gives approximately $\tan\theta_H\propto1/T^2$ and
$\rho_{xx}\propto T$ \cite{Yakovenko98a}, we conclude
that$\Delta\rho_{xx}/\rho_{xx}\propto1/T^4$, in agreement with the
experiment \cite{Ong95}.  Kohler's rule, which states that, in a
single-relaxation-time model, $\Delta\rho_{xx}(H,T)/\rho_{xx}(0,T)$ is
a function of $H/\rho_{xx}(0,T)$, is strongly violated in cuprates
\cite{Ong95}.  In a weak magnetic field, where
$\Delta\rho_{xx}/\rho_{xx}\propto H^2$, Kohler's rule requires that
$\Delta\rho_{xx}(H,T)\rho_{xx}(0,T)$ should not depend on temperature.
In Fig.\ \ref{fig:Kohler}, we show
$\Delta\rho_{xx}(H,T)\rho_{xx}(0,T)$ normalized to
$\rho_{xx}^2(0,T=\mbox{95 K})$ as a function of temperature for
different values of $\kappa$ at $H=9$ T.  We observe a strong
temperature dependence $\Delta\rho_{xx}\rho_{xx}\propto1/T^2$, which
demonstrates that Kohler's rule is violated in our model because of
the different temperature dependences of $\tau_1$ and $\tau_2$ in Eq.\
(\ref{eq:T12}).

\begin{figure}
\centerline{\psfig{file=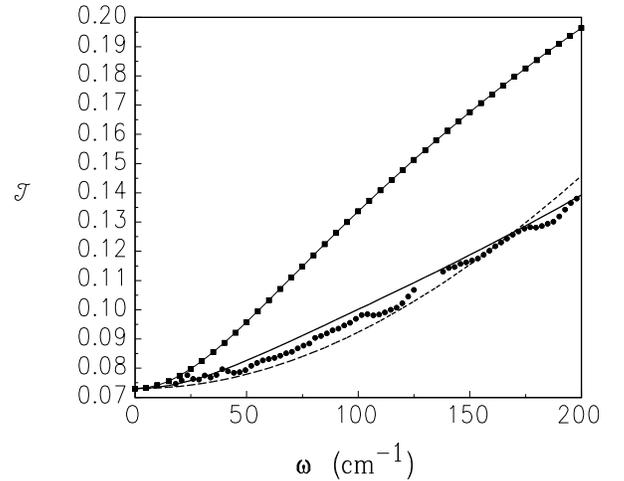,width=\linewidth,angle=-90}}
\caption{Frequency dependence of the transmittance ${\cal T}(\omega)$
  through a thin film of ${\rm YBa_2Cu_3O_7}$ in zero magnetic field.
  Dots: experimental data from Ref.\ \protect\cite{Drew96}; solid and
  dashed lines: fits to the additive and multiplicative two-$\tau$
  models of Ref.\ \protect\cite{Yakovenko98a}; solid line with
  squares: calculation for the cold spots model of Ref.\ 
  \protect\cite{Ioffe98}.}
\label{fig:Ioffe-Millis}
\end{figure}

We now briefly compare the cold spots models of Refs.\
\cite{Yakovenko98a} and \cite{Ioffe98}.  In the latter model,
transport properties are determined by the relaxation time at the
center of a cold spot, $\tau(\bar k_t)$, denoted as $\tau_{\rm FL}$ in
Ref.\ \cite{Ioffe98} and roughly corresponding to our $\tau_2$.  The
second relaxation time $\tau_1$ does not appear explicitly in the
model \cite{Ioffe98}, so only one relaxation time $\tau_{\rm FL}$
controls frequency dependence of transport properties.  The expression
for diagonal conductivity in the model \cite{Ioffe98},
\begin{equation}
  \sigma_{xx}(\omega)\propto 1/\sqrt{1-i\omega\tau_{\rm FL}},
\label{eq:xx_omega}
\end{equation}
fits the experimental data of Ref.\ \cite{Orenstein90} at $T=200$ K.
The expression for the Hall cotangent in the model \cite{Ioffe98},
\begin{equation}
  \cot\theta_H(\omega)\propto 1-i\omega\tau_{\rm FL},
\label{eq:xy_omega}
\end{equation}
fits the experimental data of Ref.\ \cite{Drew96} at $T=95$ K with
$1/\tau_{\rm FL}=54\;{\rm cm}^{-1}=7$ meV \cite{Yakovenko98a}.
Substituting this value of $1/\tau_{\rm FL}$ into Eq.\
(\ref{eq:xx_omega}), we calculate $\sigma_{xx}(\omega)$ and show the
result in Fig.\ \ref{fig:Ioffe-Millis}.  In this figure, the solid
line with squares represents frequency dependence of transmittance
through a thin film of ${\rm YBa_2Cu_3O_7}$ (which is determined by
$\sigma_{xx}(\omega)$) for the model \cite{Ioffe98}.  Dots represent
the experimental data of Ref.\ \cite{Drew96}; the solid and dashed
lines show fits for the additive and multiplicative two-$\tau$ models
studied in Ref.\ \cite{Yakovenko98a}.  The overall coefficient in Eq.\
(\ref{eq:xx_omega}) has been selected to fit the experimental value at
$\omega=0$.  We observe that the curve for the model \cite{Ioffe98}
does not match the experimental data of Ref.\ \cite{Drew96}.  The
reason for this failure is that fitting both $\sigma_{xx}(\omega)$ and
$\sigma_{xy}(\omega)$ requires two different frequency scales, such as
$1/\tau_1$ and $1/\tau_2$ \cite{Yakovenko98a}, whereas the model
\cite{Ioffe98} has only one frequency scale $1/\tau_{\rm FL}$.

In conclusion, we have calculated the in-plane magnetoresistance
$\Delta\rho_{xx}/\rho_{xx}$ of YBa$_2$Cu$_3$O$_7$ in a magnetic field
applied perpendicular to the $\rm CuO_2$ planes for the additive
two-$\tau$ model of Ref.\ \cite{Yakovenko98a}.  We find that Kohler's
rule is strongly violated, but the ratio
$\Delta\rho_{xx}/\rho_{xx}\tan^2\theta_H$ is approximately
temperature-independent and, for the model Fermi surface employed, is
approximately equal to 5.5.  The theoretical results agree
qualitatively and, within an order of magnitude, quantitatively with
the experiment.

We are grateful to I.~I.~Mazin for providing the band-structure data
of YBa$_2$Cu$_3$O$_7$ and to A.~J.~Millis for useful discussions.

\end{document}